\documentclass[conference]{IEEEtran}
\IEEEoverridecommandlockouts
\usepackage[letterpaper, left=0.625in, right=0.625in, bottom=1in, top=0.71in]{geometry}

\usepackage{cite}
\usepackage{amsmath,amssymb,amsfonts}
\usepackage[ruled]{algorithm2e}
\usepackage{graphicx}
\usepackage{textcomp}
\usepackage{xcolor}
\usepackage{cleveref}
\usepackage{bm}
\usepackage{comment}
\usepackage{stfloats}

\usepackage{siunitx}
\DeclareSIUnit\BL{\giga\bit\kilo\meter\per\second}
\DeclareSIUnit\FL{\decibel\per\kilo\meter}
\DeclareSIUnit\CD{\pico\second\per\nano\meter\per\kilo\meter}
\DeclareSIUnit\baud{Bd}

\usepackage{mleftright}
\DeclareMathOperator{\E}{\mathbb{E}}

\newcommand{\elr}{\epsilon_{\mathrm{lr}}} %
\newcommand{\Nos}{N_{\mathrm{os}}}    %
\newcommand{\Nti}{N_{\mathrm{TI}}}    %
\newcommand{\Nd}{N_{\mathrm{d}}}    %
\newcommand{\Nch}{N_{\mathrm{ch}}}    %
\newcommand{\Nparam}{N_{\mathrm{param}}}    %
\newcommand{\Fma}{F_{\mathrm{ma}}}    %
\newcommand{\varN}{\sigma^2_{\mathrm{n}}} %
\newcommand{\varRefi}{\sigma_\mathrm{ref}^{2}\mlr{i}} %
\newcommand{\Lf}{L_{\mathrm{fiber}}} %
\newcommand\snr{{\mathsf{SNR}}}

\newcommand{\mlr}[1]{\mleft(#1\mright)}
\newcommand{\mlrb}[1]{\mleft\{#1\mright\}}

\newcommand{\multiNormDis}[3]{\mathcal{N}_{#3}\mlr{#1,\,#2}}

\usepackage[nohyperlinks,nolist]{acronym}
\acrodef{ai}[AI]{artificial intelligence}
\acrodef{ask}[ASK]{amplitude-shift keying}
\acrodef{awgn}[AWGN]{additive white Gaussian noise}

\acrodef{bce}[BCE]{binary cross-entropy}
\acrodef{bpsk}[BPSK]{binary phase-shift keying}

\acrodef{cd}[CD]{chromatic dispersion}
\acrodef{cir}[CIR]{combined impulse response}
\acrodef{cma}[CMA]{constant modulus algorithm}
\acrodef{cnn}[CNN]{convolutional neural network}
\acrodef{ctp}[CTP]{channel transition probability}
\acrodefplural{ctp}[CTPs]{channel transition probabilities}

\acrodef{dnn}[DNN]{deep neural network}
\acrodef{dsp}[DSP]{digital signal processing}

\acrodef{fc}[FC]{fully connected}
\acrodef{fec}[FEC]{forward error correction}
\acrodef{fir}[FIR]{finite impulse response}

\acrodef{elu}[ELU]{exponential linear unit}

\acrodef{gan}[GAN]{generative adversarial network}
\acrodef{hwa}[HWA]{historical weight averaging}

\acrodef{imdd}[IM/DD]{intensity modulation with direct detection}
\acrodef{iot}[IoT]{Internet of things}
\acrodef{isi}[ISI]{inter-symbol interference}

\acrodef{lms}[LMS]{least-mean-square}

\acrodef{mlsd}[MLSD]{maximum-likelihood sequence detection}
\acrodef{mse}[MSE]{mean-squared-error}
\acrodef{ma}[MA]{moving average}
\acrodef{map}[MAP]{maximum a posteriori}

\acrodef{nn}[NN]{neural network}
\acrodef{ns}[NS]{non-saturating}

\acrodef{pam}[PAM]{pulse-amplitude modulation}
\acrodef{pdf}[pdf]{probability density function}
\acrodef{pmf}[pmf]{probability mass function}

\acrodef{rc}[RC]{raised-cosine}
\acrodef{relu}[ReLU]{rectified linear unit}

\acrodef{ser}[SER]{symbol error rate}
\acrodef{sld}[SLD]{square-law detection}
\acrodef{snr}[SNR]{signal-to-noise ratio}
\acrodef{sps}[sps]{samples per symbol}
\acrodef{ssmf}[SSMF]{standard single-mode fiber}

\acrodef{ti}[TI]{training iteration}

\acrodef{vnle}[VNLE]{Volterra-based nonlinear equalizer}

\begin{document}

\title{Blind and Channel-agnostic Equalization\\Using Adversarial Networks
\thanks{This work was carried out in the framework of the CELTIC-NEXT project AI-NET-ANTILLAS (C2019/3-3) and was funded by the German Federal Ministry of Education and Research (BMBF) under grant agreements 16KIS1316 and 16KIS1317 as well as under grant 16KISK004 (Open6GHuB).}
}

\author{
\IEEEauthorblockN{Vincent Lauinger${}^\star$, Manuel Hoffmann${}^\star$, Jonas Ney${}^\ddagger$, Norbert Wehn${}^\ddagger$, and Laurent Schmalen${}^\star$}
\IEEEauthorblockA{${}^\star$\textit{Communications Engineering Lab (CEL)}, \textit{Karlsruhe Institute of Technology (KIT)}, Germany \\
\{\texttt{vincent.lauinger}, \texttt{laurent.schmalen}\}\texttt{@kit.edu} \\
${}^\ddagger$\textit{Microelectronic Systems Design (EMS)}, \textit{University of Kaiserslautern}, Germany \\
\{\texttt{ney}, \texttt{wehn}\}\texttt{@eit.uni-kl.de}}
}

\maketitle

\begin{abstract}
Due to the rapid development of autonomous driving, the Internet of Things and streaming services, modern communication systems have to cope with varying channel conditions and a steadily rising number of users and devices. This, and the still rising bandwidth demands, can only be met by intelligent network automation, which requires highly flexible and blind transceiver algorithms.
To tackle those challenges, we propose a novel adaptive equalization scheme, which exploits the prosperous advances in deep learning by training an equalizer with an adversarial network. The learning is only based on the statistics of the transmit signal, so it is \emph{blind} regarding the actual transmit symbols and \emph{agnostic} to the channel model. The proposed approach is independent of the equalizer topology and enables the application of powerful neural network based equalizers.
In this work, we prove this concept in simulations of different---both linear and nonlinear---transmission channels and demonstrate the capability of the proposed blind learning scheme to approach the performance of non-blind equalizers.
Furthermore, we provide a theoretical perspective and highlight the challenges of the approach.

\end{abstract}


\acresetall

\section{Introduction} \label{Sec:intro}

The digital transformation and the advent of video streaming platforms brought up a strong demand for high-speed and highly flexible communication systems. 
Along with the rapid development of autonomous driving, the Internet of Things and Industry 4.0, intelligent automation becomes highly important for modern communication networks to adapt quickly to varying channel conditions. %
This adaptation should not only be performed quickly but also fully autonomously to enable the design of fast, low-cost and low-latency transmission systems for a broad range of applications, where no further expert knowledge or extensive parameter searches are required during run-time. This requires high flexibility of the signal processing algorithms;
however, state-of-the-art algorithms usually assume certain channel models or transmission characteristics in their training objective. The \ac{cma}~\cite{godard_cma} and its derivatives, e.g., assume symmetric single-level modulation formats. Other algorithms assume specific channels, e.g., the \ac{awgn} channel or the nonlinear optical channel described by the Manakov equation.
Channel-agnostic algorithms, however, can also be used if the channel model is unknown or if it deviates significantly from the real channel.
Furthermore, modern state-of-the-art transmission systems often rely on data-aided (or pilot-based) signal processing algorithms, however, the resulting overhead limits the achievable net bit rate significantly. %
Hence, there is a strong demand on blind and flexible transceiver algorithms.

In this paper, we propose a novel blind and channel-agnostic approach to train an equalizer with an adversarial network, %
while, to the best of our knowledge, there is no blind state-of-the-art equalizer which is fully agnostic to the channel model.
Our approach is based on \acp{gan} \cite{goodfellow_gan}, which represent an unsupervised learning concept based on two \acp{nn}, the generator and the discriminator, where both are trained jointly in a zero-sum game. %
The concept is commonly used in image processing and computer vision, e.g., to increase the amount of training data~\cite{antoniou_data_aug_gan} or for generating realistic artificial photographs~\cite{Karras_progressivGAN, brock_largeScaleGAN}. In recent years, multiple works focused on applying the \ac{gan} concept to communication systems. The most common use-case %
is autonomous channel modeling, where the \ac{gan} is trained on raw measurements to obtain an appropriate and differentiable channel model~\cite{yang_GANmodelling, sun_GAN_viterbi, karanov2020gan, yang_fast}. 
For instance, a \ac{gan} is used to predict unknown \aclp{ctp}, which are further utilized by the Viterbi algorithm for channel-independent \ac{mlsd}~\cite{sun_GAN_viterbi}. 
However, the approach still faces the common issues of \ac{mlsd}, e.g., exponentially-increasing complexity with increasing channel memory, which makes it impractical for many realistic communication channels.

Contrarily, we focus on a different approach which is more similar to%
~\cite{nguyen_GAN}, where an error correcting decoder is trained with adversarial networks. Instead of using the generator to create new data samples from noise as input, as in most \ac{gan} applications for channel modelling, \cite{nguyen_GAN} feeds the generator with the received noisy codewords. The generator serves as decoder to recover the transmitted codewords and is trained together with the adversarial discriminator, which tries to distinguish between the decoder output and valid reference codewords. %
In a similar way, instead of recovering noisy codewords, we focus on equalizing a received sequence of symbols, which is not only corrupted by noise but also distorted by potential nonlinearities and effects with memory, e.g., \ac{isi}. 
However, the distortions cause our problem to be much more demanding than the recovery of noisy codewords. 
In this work, we propose a concept to utilize the novel \ac{gan} approach for blind and channel-agnostic equalization, and prove it in simulations of these demanding conditions. 
Compared to \cite{nguyen_GAN}, we use the generator as equalizer (instead of decoder), feed it with the received sequence and train it with an adversarial discriminator based on reference sequences.
In fact, we only require knowledge of the transmitted sequences' statistics to draw new reference sequences at the receiver, but do not need any supervision or channel model in training. 
Hence, the novel training scheme is blind, channel-agnostic and independent of the equalizer topology.
We demonstrate its high flexibility %
and introduce concepts to improve training along with appropriate topologies based on both linear equalizers and \acp{nn}. 
Furthermore, we analyze the approach from a theoretical perspective and show that it is potentially able to outperform state-of-the-art equalizers.

\section{Concept of Generative Adversarial Networks}
\label{Sec:GAN_concept}

In this work, we exploit the concept of \acp{gan} to train an equalizer (generator) using an adversarial discriminator.

A \ac{gan} consists of two individual (parametric) functions---generator $G$ and discriminator $D$---which are usually represented by \acp{nn} and trained in an adversarial manner~\cite{goodfellow_gan}. Precisely, they play a minimax-game, i.e., optimize the value function 
\begin{align*}
    \begin{split}
        \min_G\,\max_D \,V(D,G) &= \E_{\bm{x}^\prime\sim p\mlr{\bm{x}}}\mlrb{\log{D\mlr{\bm{x}^\prime}}} \\
    &\qquad+ \E_{\bm{y}\sim p\mlr{\bm{y}}}\mlrb{\log\mlr{1-D\mlr{G\mlr{\bm{y}}}}}
    \end{split}  %
\end{align*}
with $\E_{\bm{y}\sim p\mlr{\bm{y}}}\mlrb{\cdot}$ being the expectation with respect to $p\mlr{\bm{y}}$ and $\bm{x}^\prime$ being reference samples drawn from the desired \ac{pdf} $p\mlr{\bm{x}}$. In simple words, $G$ tries to ``fool'' the discriminator by generating an output---usually out of noise as input---which resembles the reference samples $\bm{x}^\prime$, while $D$---which outputs a scalar representing the probability of its input being a reference sample---tries to discriminate correctly between generator output $G\mlr{\bm{y}}$ and reference samples $\bm{x}^\prime$. 

Goodfellow~et~al. showed that the minimax-game's global optimum is at $p\mlr{\bm{z}} = p\mlr{\bm{x}}$ (with $\bm{z}=G\mlr{\bm{y}}$) and that their proposed training \cite[Algorithm~1]{goodfellow_gan} converges to this optimum, if ``$G$ and $D$ have enough capacity and, at each step of Algorithm~1, the discriminator is allowed to reach its optimum given $G$''~\cite{goodfellow_gan}. 
Importantly, $G$ and $D$ are updated blindly (without supervision) using any gradient-based learning rule%
, e.g.~\cite{kingma2014adam}, with the aid of the reference symbols, which can be generated at the receiver.

\section{GAN-based Equalization} \label{Sec:GAN}
\begin{figure}[tb]
	\centerline{\includegraphics{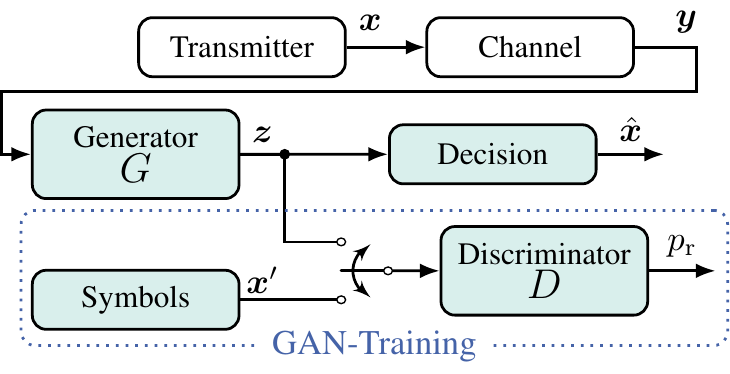}}
	\caption{Structure of the proposed \ac{gan}-based blind equalization approach. Data symbols $\bm{x}$ are transmitted over an unknown communication channel to a receiver. Only the colored components belong to the receiver and the components in the dotted box are specific to the \ac{gan} approach. The generator $G$ is the equalizer, $\bm{x}^\prime$ are reference symbols (generated at the receiver) and $\hat{\bm{x}}$ are the recovered symbols.}
	\label{fig:gan_approach}
\end{figure}

In our novel approach, we utilize the \ac{gan} for blind channel-agnostic equalization by considering the equalizer as generator $G$ and introducing an additional discriminator $D$ for adversarial learning, as depicted in Fig.~\ref{fig:gan_approach}. 

Since the proposed equalizer is channel-agnostic, we assume a general communication system where the transmitted vector $\bm{x}$, consisting of symbols $x_i\in\mathcal{A}$ from an alphabet $\mathcal{A} = \mlrb{A_1 \ldots A_M}$, is distorted by an unknown channel and results in a received vector $\bm{y}$.

Instead of inputting noise to the generator, as in most \ac{gan} implementations, we pass the received vector $\bm{y}$ to the generator to operate it like a conventional equalizer.
The proposed approach allows various linear and nonlinear equalizer and discriminator topologies, so it is not limited to specific \acp{nn}; however, it has to be highlighted that the topologies determine 
the equalizer output $\bm{z}$ as well as the family of potential
$p\mlr{\bm{z}}$, which can be generated by the equalizer. Hence, the topologies have to be chosen carefully.

\subsection{Training} \label{Sec:setup_training}
\newcommand\mycommfont[1]{\footnotesize\ttfamily\textcolor{blue}{#1}}
\SetCommentSty{mycommfont}
\DontPrintSemicolon

\SetKwFor{For}{for (}{)}{}

\begin{algorithm}
\normalsize 
\vspace*{0.1cm}
\label{Alg:training_flow}
\caption{Training flow of the \ac{gan}-equalizer}\label{alg:training}

\SetNoFillComment

$\Nti$: Number of \acsp*{ti}; \hspace{8pt}$N$: Number of symbols per \acs*{ti}\;
$\Nd$: Number of discriminator updates per \acs*{ti}\;
$\multiNormDis{\mu}{\sigma^{2}}{N}$: $N$-dimensional normal distribution with each dimension's mean $\mu$ and variance $\sigma^{2}$\;
$\gamma_\text{ch}\mlr{\bm{x}}$: Function representing the channel distortions\;
$\varRefi$: Function returning the reference's variance at the \acs*{ti} $i$

\BlankLine
\For{$i=0$;\ $i<\Nti$;\ $i\text{++}$}{
	    \tcc{Train Discriminator}
	    \For{$j=0$;\ $j<\Nd$;\ $j\text{++}$}{
	    \tcc{Sample reference symbols}
	    $ \bm{x}' \sim P\mlr{\bm{x}} + \multiNormDis{0}{\varRefi}{N}$\;
	    \tcc{Sample channel output}
	    $ \bm{y} \sim \gamma_\text{ch}\mlr{\bm{x}} + \multiNormDis{0}{\varN}{N}$ with $\bm{x} \sim p\mlr{\bm{x}}$\;
	    \tcc{Generate fake symbols}
	    $\bm{z} = G(\bm{y})$\;
	    \tcc{Get predictions of D to calc loss}
	    $\text{loss}_\mathrm{D} = \text{BCE}(D(\bm{x}'), 1) + \text{BCE}(D(\bm{z}), 0) + \text{HWA}_\mathrm{D}$ \;
	    update parameters of $D$ based on $\text{loss}_\mathrm{D}$\;
    }
    \tcc{Train Generator}
    \tcc{Get prediction for $\bm{z}$ and calc loss}
    $\text{loss}_\mathrm{G} = \text{BCE}(D(\bm{z}), 1) + \text{HWA}_\mathrm{G}$\;
    update parameters of $G$ based on $\text{loss}_\mathrm{G}$\;
}
\end{algorithm}
In order to train the generator and discriminator, we make use of the \ac{ns}-\ac{gan} approach described in~\cite{goodfellow_gan, salimans_improved}, where the generator is trained by maximizing $\log{D\mlr{G\mlr{\bm{y}}}}$ instead of minimizing $\log\mlr{1-D\mlr{G\mlr{\bm{y}}}}$. This prevents the saturation of the generator loss function early in learning (when $G$ is poor, so $D$ can easily distinguish between $\bm{z}$ and $\bm{x}^\prime$) but has the same fixed point of $G$'s and $D$'s dynamics~\cite{goodfellow_gan}.
\begin{figure}[!tb]
	\centerline{\includegraphics{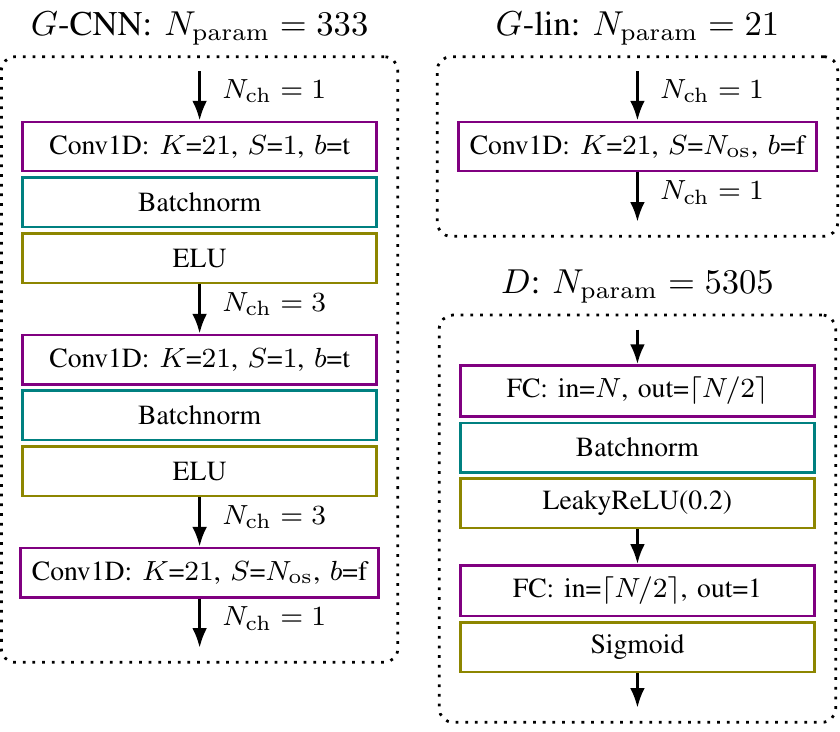}}
	\caption{Topology of the generator $G$ (CNN and lin) and discriminator $D$ with kernel size $K$, stride $S$, oversampling factor $\Nos$, number of channels $\Nch$, number of trainable parameters $\Nparam$, and the  bias $b$ being \emph{true} (t) or \emph{false}~(f).}
	\label{fig:network_topology}
\end{figure}
Additionally, we tackle the same issue by blurring $\bm{x}'$ with Gaussian noise, whose variance $\varRefi$ decreases with the \acp{ti}. %
As suggested in~\cite{salimans_improved}, 
we apply \acfi{hwa} for both $G$ and $D$ to increase the training stability. Precisely, we penalize parameter updates with a large difference to the corresponding parameters' \ac{ma} of the past 200 \acp{ti}.

The complete training flow is described in Algorithm~\ref{Alg:training_flow}.
We use \ac{bce} loss for both the generator and the discriminator, and have two separate optimizers for $G$ and $D$. Thus, both can be trained with individual update rules and learning rates $\elr$. In this work, both use the Adam algorithm~\cite{kingma2014adam}. %
We train the models with $\Nd=2$, $N=100$ symbols per \ac{ti} and apply a learning rate scheduler, which multiplies the learning rate by $0.3$ after predefined \acp{ti} steps.

\subsection{Topology} \label{Sec:setup_topology}
The structures of the \acp{nn} used to achieve the results presented in Sec.~\ref{Sec:simu} are illustrated in~Fig.~\ref{fig:network_topology}; however, we want to highlight that the approach itself is topology-independent and that we do not claim that the proposed structures are optimal. Yet, we performed an extensive design space exploration to find topologies of low complexity, which achieve satisfying results in our simulations. We leave further topology optimization for future research.

Precisely, we implemented two structures for the generator as depicted in Fig.~\ref{fig:network_topology}: $G$-\acs*{cnn} is based on a \ac{cnn} and $G$-lin is based on a linear \ac{fir} filter. %
$G$-\ac{cnn} is composed of three one-dimensional convolutional layers (conv1D) with, except for the last layer, batchnorm~\cite{Ioffe_batchnorm} and \ac{elu} activation. The intermediate layers expand to three channels increasing the \ac{cnn}'s expressive power and the last layer downsamples by a stride $S=\Nos$, where $\Nos$ is the oversampling factor in \ac{sps}. 
$G$-lin corresponds to $G$-\ac{cnn}'s last layer, but only has one input and output channel.
We use  a uniform \emph{Glorot} initialization in the first layer and \emph{Dirac} in the remaining layers (as well as for $G$-lin), since we expect that this fits best to the final weight values. %
Again, we propose the optimization of the initializations as subject of future research.

The discriminator receives a sequence as input but has to output a scalar (probability), so we implemented this funnel-like shape with a two-layer \ac{fc} \ac{nn} as depicted in Fig.~\ref{fig:network_topology}. %

\subsection{Theoretical Perspective} \label{Sec:GAN_it}

\begin{figure} [tb] %
	\begin{center}
		\centerline{\includegraphics{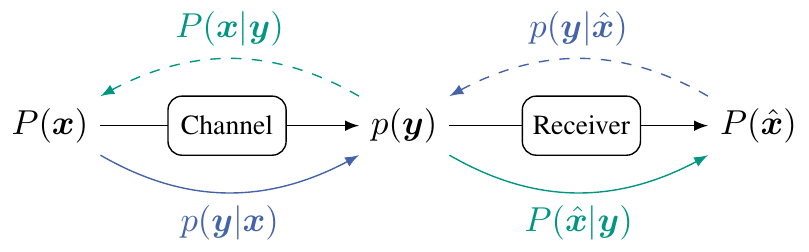}}
		\caption{Simple system model and its corresponding \acp{pdf} and \acp{pmf}: transmitted vector $\bm{x}$, received vector $\bm{y}$ and recovered vector $\hat{\bm{x}}$ %
		are realizations of random variable sequences $\lbrace X_k\rbrace$, $\lbrace Y_k\rbrace$ and $\lbrace \hat{X}_k\rbrace$. %
		}
		\label{fig:sys_mod}
	\end{center}
\end{figure}

Additionally to the rather practical deep learning perspective in the previous sections, in the following, we analyse the approach from a theoretical point of view.
Therefore, we depict a general communication system %
as in Fig.~\ref{fig:sys_mod}.
The transmitted vector $\bm{x}$ is a realization of the random variable sequence $\mlrb{X_k}\sim P\mlr{\bm{x}}$, where $P\mlr{\bm{x}}$ is the corresponding \acf{pmf}. The channel distorts $\bm{x}$ and, thus, transforms the random variable sequence $\mlrb{X_k}$. If there is superimposed continuously-distributed noise, e.g., \ac{awgn}, the received symbols' random variable sequence follows a \ac{pdf}, i.e., $\mlrb{Y_k}\sim p\mlr{\bm{y}}$.

The optimum decision is based on the \ac{map} probability~\cite[Ch.~4.1]{proakis2008digital}
\begin{align*}
	P\mlr{\bm{x}|\bm{y}} &= \frac{p\mlr{\bm{y}|\bm{x}} \cdot P\mlr{\bm{x}}}{p\mlr{\bm{y}}} \ ,
\end{align*}
where the likelihood $p\mlr{\bm{y}|\bm{x}}$ describes the channel. If $p\mlr{\bm{y}|\bm{x}}$ is unknown, $P\mlr{\bm{x}|\bm{y}}$ is intractable to compute. Hence, Fig.~\ref{fig:sys_mod} reveals that we have to find a receiver, which equalizes $\bm{y}$ such that
\begin{align*}
	\frac{p\mlr{\bm{y}| \hat{\bm{x}}} \cdot P\mlr{\hat{\bm{x}}}}{p\mlr{\bm{y}}} = P\mlr{\hat{\bm{x}}| \bm{y}} &\overset{!}{=} P\mlr{\bm{x}| \bm{y}} = \frac{p\mlr{\bm{y}| \bm{x}} \cdot P\mlr{\bm{x}}}{p\mlr{\bm{y}}} \; ,
\end{align*}
where $\hat{\bm{x}}$ is the output after equalizer and demapper. %
One possible solution for this problem is
\begin{align}
	p\mlr{\bm{y}| \hat{\bm{x}}} \overset{(i)}{=} p\mlr{\bm{y}| \bm{x}} \quad&\text{and} \quad  P\mlr{\hat{\bm{x}}} \overset{(ii)}{=} P\mlr{\bm{x}} \; . \label{IT_requirements}
\end{align}
While $(ii)$ corresponds to the GAN's training objective (if the demapper is included into the approach), the satisfaction of requirement $(i)$ is the crucial point. %

The analysis shows that future research has to focus especially on the satisfaction of requirement $(i)$ by adequate system design, but it also reveals that the proposed blind and channel-agnostic GAN based approach is potentially able to converge to the optimal \ac{map} solution.

\section{Simulations} \label{Sec:simu}
We evaluate our approach by simulations of different communication channels and present the results in this section. %

\subsection{Channel Model} \label{Sec:simu_channel}
\begin{figure}[tbp]
	\centerline{\includegraphics{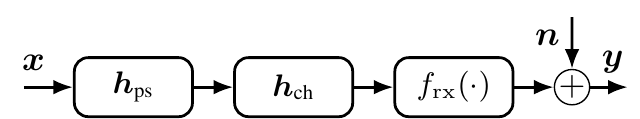}}
	\caption{Generalized channel model where the vector $\bm{x}$ contains the transmitted symbols, $\bm{h}_{\mathrm{ps}}$ and $\bm{h}_{\mathrm{ch}}$ are the linear impulse responses of the pulse shaping filter and the channel, $f_{\mathrm{rx}}$ describes a potential non-linear distortion (which is applied element-wise), $\bm{n}$ is \ac{awgn}, and $\bm{y}$ is the received vector.}
	\label{fig:channel}
\end{figure}

We demonstrate the flexibility of our approach by evaluating it under different, both linear and nonlinear, communication channels. All simulated channels can be described by the generalized model in Fig.~\ref{fig:channel}, where the transmitted symbols, represented by the vector $\bm{x}$, are convolved with a \ac{rc} pulse shaping filter $\bm{h}_\mathrm{ps}$ (with roll-off $\rho_\mathrm{rc}=0.25$) and a linear channel impulse response $\bm{h}_\mathrm{ch}$. Specific receiver characteristics can be described by a function $f_\mathrm{rx}\mlr{\cdot}$, which is element-wise applied to a vector. Finally, the received vector $\bm{y}$ is superimposed by a Gaussian noise vector $\bm{n}$. Since oversampling is essential to real systems, we run all simulations at an oversampling rate of $\Nos=2$~\ac{sps}.

Precisely, we simulate the bad-quality linear telephone channel described in \cite[Ch.~9.4-3]{proakis2008digital}, known as \emph{Proakis-B} (ProB), with equivalent discrete-time impulse response
\begin{align*}
	\bm{h}_{\mathrm{ch,\, ProB}} &= [0.407, \ 0.815, \ 0.407] \; ,
\end{align*}
which we upsample to $\Nos=2$~\ac{sps} by interleaving a zero between each tap. The pulse shaping filter then interpolates both the channel and the transmitted vector (which is also upsampled by inlerleaving zeros). 
For the linear channel, we use a real-valued \ac{bpsk} modulation and have $f_\mathrm{rx}\mlr{\tilde{\bm{x}}}=\tilde{\bm{x}}$.

Additionally, we simulate a nonlinear, dispersive optical channel with \ac{imdd} and \ac{pam} as described in \cite{plabst_wiener}. The \ac{sld} at the receiver distorts the signal nonlinearly and can be modeled by
$\tilde{\bm{y}} = f_\mathrm{rx}\mlr{\tilde{\bm{x}}}$ with $\tilde{y}_i=|\tilde{x}_i|^2$. %
Linear channel distortions are caused by \ac{cd}, which can be described (including attenuation) by its frequency response
\begin{equation*}
	H_\mathrm{cd}\left( \Lf, f \right) = \exp\left(- \frac{1}{2} \alpha \Lf \, + \, \j 2 \pi^2 \beta_2 f^2 \Lf\right)  \label{eq:cd_freqency_response}
\end{equation*}
where $\Lf$ is the fiber length, $\beta_2=-\frac{\lambda^2}{2\pi \mathrm{c}} D_\mathrm{cd}$ is defined by the wavelength $\lambda$, the speed of light $\mathrm{c}$ and the fiber's dispersion coefficient $D_\mathrm{cd}$; and $\alpha$ is the fiber attenuation.
This work considers  C-band transmission at $\lambda=\SI{1550}{\nano\meter}$ over a \ac{ssmf}, i.e., $D_\mathrm{cd} = \SI{17}{\CD}$ and $\alpha \triangleq \SI{0.2}{\FL}$.

As both the pulse shaping filter and \ac{cd} filter are linear and time-invariant, we can define the \acl*{cir} 
\begin{equation*}
    \bm{h}_\text{cir} = \bm{h}_\text{ps}\ast \bm{h}_\text{ch} = \mathcal{F}^{-1}\mlrb{\bm{H}_\text{ps}\cdot \bm{H}_\text{cd}} 
\end{equation*}
where $\mathcal{F}^{-1}\mlrb{\cdot}$ is the inverse Fourier transform and $\bm{H}_\text{ps}$ is the frequency response of the pulse shaping filter. 

Following \cite{plabst_wiener},~\cite[ch.~4]{Agr10}, we assume to have no optical amplifiers, so the dominant noise source is the thermal noise in the electronic domain after the \ac{sld}, which can be modeled as \ac{awgn} with zero-mean and the variance $\varN$. For the simulation, we fix the \ac{snr} to $\SI{20}{dB}$ (PAM-2) and $\SI{26}{dB}$ (PAM-4). 
At the receiver, we conduct hard decision based on the minimum Euclidean distance and compensate for shifts, scaling-offsets, inversions and permutations, which can occur at blind equalization. In real systems, this can be done by transmitting a header or testing the possible combinations with the \ac{fec} decoder.

\subsection{Reference Algorithms} \label{Sec:simu_reference}
As baseline, we train the same equalizer topologies, $G$-\ac{cnn} and $G$-lin, as used for the proposed blind channel-agnostic \ac{gan}-approach with the \ac{mse} loss in a supervised manner, i.e., with knowledge of the transmitted symbols, and call it ``sup-CNN'' and ``sup-lin'' in the following. 
Additionally, for the \ac{imdd} channel, we evaluate the received vector without any equalization---further denoted as ``woEq''---as second baseline. Furthermore, we implement an equalizer topology based on a third-order \emph{Volterra} kernel~\cite{stojanovic2017volterra} with memory $F=[35, 17, 9]$, which corresponds to 354 trainable parameters. We also train it with the \ac{mse} loss in a supervised manner and call it ``sup-Volt''.  
For the linear channels, we implement the basic data-aided \ac{lms} algorithm to update an \ac{fir} filter of length $F=21$ based on gradient-descent in a supervised manner \cite{proakis2008digital}. 

\subsection{Acquisition Scheme} \label{Sec:simu_acqu}
Since we want to demonstrate the concept by evaluating its potential performance, we display the best averaged performance of each algorithm. Although the approach faces training instabilities, we can demonstrate its performance by both averaging over multiple simulation runs and estimates per run.
Therefore, we estimate the \ac{ser} after every $\Nti/100$ \acp{ti} over 10,000 symbols, perform an \ac{ma} with filter length $\Fma=10$ over the sequence of \ac{ser} estimates and choose the minimum of the $(100-\Fma+1)$ values. Furthermore, we conduct five (unless stated differently) independent simulation runs with the same hyperparameter settings but randomly initialized weights to analyze the algorithm's stability. 
We define a run as \emph{successful} if the difference between its minimum \ac{ser} after \ac{ma} to the smallest of all runs is less than a predefined threshold $\mathrm{SER}_\mathrm{th}$. Precisely, we heuristically set $\mathrm{SER}_\mathrm{th} = 0.07$ to include as much variance due to gradient noise and local minima as possible by filtering runs which did not converge properly at all. 
The depicted \emph{mean SER} is averaged over all successful runs and the amount of successful runs is depicted as small number next to the corresponding point in the result figures. If the \emph{minimum SER} is depicted as well, the value corresponds to the minimum value before the \ac{ma} filtering and over all runs.

\subsection{Numerical Results} \label{Sec:simu_results}

\begin{figure}[tb]
	\centering
	\includegraphics{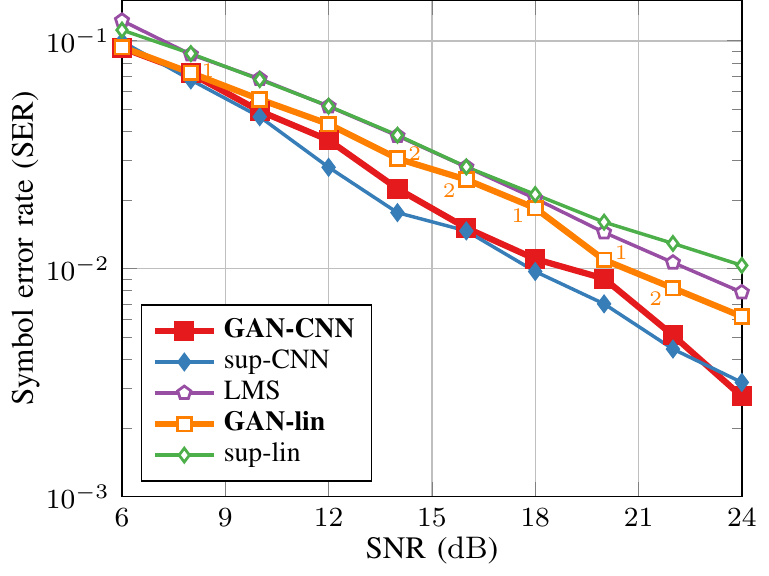}
	\caption{Results for BPSK transmission over the \emph{ProB} \ac{awgn} channel with \ac{isi} --- GAN: $\elr|_\mathrm{G}=0.001$, $\elr|_\mathrm{D}=0.0005$; LMS: $\elr=0.01$; supervised (sup): $\elr|_\mathrm{CNN}=0.005$, $\elr|_\mathrm{lin}=0.001$. If applicable, a small digit next to a marker depicts the number of non-converged runs per 5 simulation runs.}
	\label{fig:BPSK_snr_nma}
\end{figure}

The results for the linear \ac{awgn} channel with \ac{isi} are depicted in Fig.~\ref{fig:BPSK_snr_nma}. As stated in \cite[Ch.~9.4-3]{proakis2008digital}, nonlinear equalizers have advantages in the \emph{ProB} channel, which has zeros in its frequency response. 
The GAN-CNN is able to use these and significantly outperforms linear (non-blind) equalizers. In fact, it even follows the track of the non-blind sup-CNN without failing to converge. Interestingly, the GAN-lin significantly outperforms the other equalizers based on the same linear filters (sup-lin and LMS) by one to two dB, so the GAN may have the better training objective as state-of-the-art linear equalizers for channels with zeros in its frequency response; however, it fails to converge in some cases.

\begin{figure}[tb]
	\centering
	\includegraphics{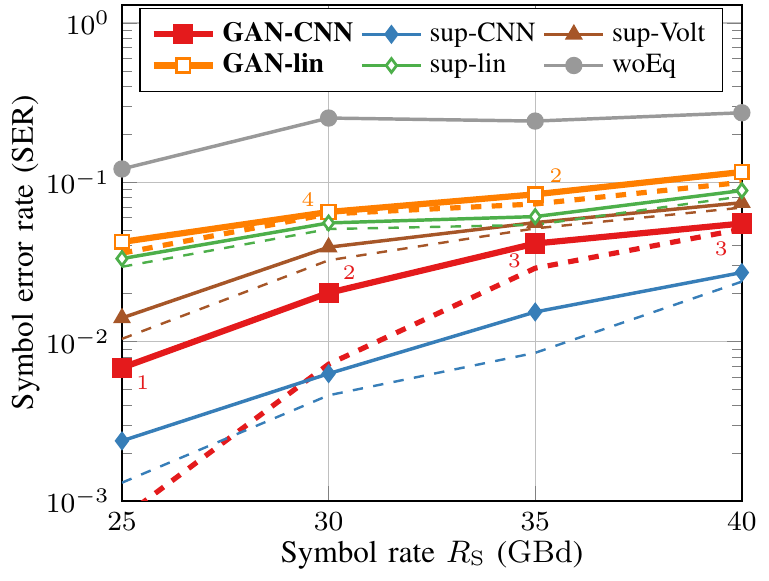}
	\caption{Results for PAM-2 transmission over an \ac{imdd} channel with $\Lf=\SI{30}{\kilo\meter}$, $\snr=\SI{20}{dB}$ --- GAN: $\elr|_\mathrm{G}=0.001$, $\elr|_\mathrm{D}=0.001$; sup: $\elr|_\mathrm{CNN}=0.0005$, $\elr|_\mathrm{lin}=0.005$, $\elr|_\mathrm{Volt}=0.007$. Depicted is the mean SER (\textbf{-----}), the minimum SER  (\textbf{- - -}), and the number of non-converged runs per 5 simulation runs.}%
	\label{fig:PAM2_f_nma}
\end{figure}

\begin{figure}[b]
	\centering
	\includegraphics{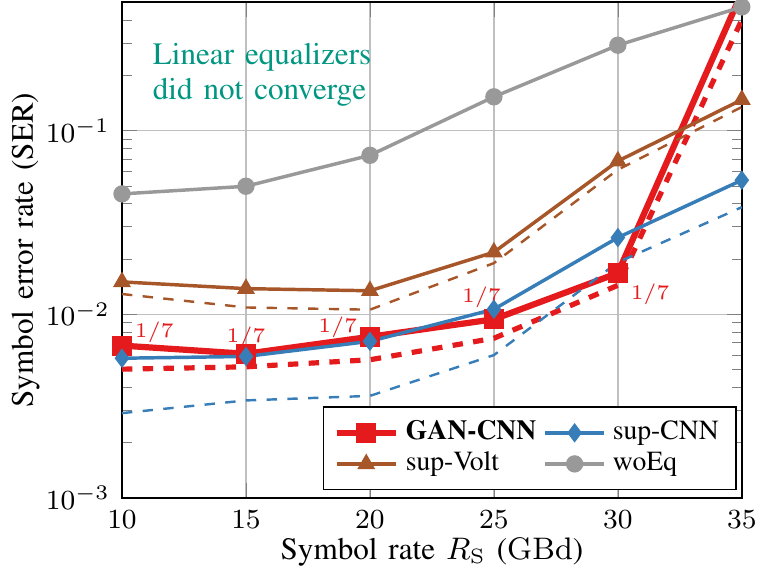}
	\caption{Results for PAM-4 transmission over an \ac{imdd} channel with $\Lf=\SI{15}{\kilo\meter}$, $\snr=\SI{26}{dB}$ --- GAN: $\elr|_\mathrm{G}=0.001$, $\elr|_\mathrm{D}=0.001$; sup: $\elr=0.002|_\mathrm{CNN}$, $\elr|_\mathrm{Volt}=0.003$. Depicted is the mean SER (\textbf{-----}), the minimum SER  (\textbf{- - -}), and the number of non-converged runs per 7 simulation runs.}
	\label{fig:PAM4_f_nma}
\end{figure}

The results for \ac{pam}-2 transmission over the \ac{imdd} channel are depicted in Fig.~\ref{fig:PAM2_f_nma} for multiple symbol rates.
The proposed GAN-CNN is following the non-blind sup-CNN's track with a significant penalty, but still outperforms the other non-blind reference algorithms; however, the convergence is not as reliable as for the linear channel. Nevertheless, the minimum \ac{ser} is similar to (or even better as) the sup-CNN's minimum \ac{ser}, which proves the concept and motivates further research on stabilizing the \ac{gan} training. In this nonlinear channel, the GAN-lin is not able to outperform the baseline non-blind linear equalizers, but converges to their performance.

The GAN-CNN is also able to converge for multi-amplitude modulation formats, as proven in Fig.~\ref{fig:PAM4_f_nma} for PAM-4 transmission. 
For symbol rates below \SI{35}{\giga\baud}, the GAN-CNN reaches the same performance or even outperforms (for \SI{30}{\giga\baud}) the non-blind sup-CNN. The non-blind Volterra (sup-Volt) equalizer faces a significant penalty to the CNN-based equalizers, although both the Volterra and CNN structures have a comparable number of optimizable parameters $\Nparam$~(354~vs.~333). For multi-amplitude formats, where the nonlinearities have a greater impact, the linear equalizers do not converge anymore.
The cause for the GAN-CNN's performance deterioration at \SI{35}{\giga\baud} is part of our ongoing investigations.

\begin{figure*}[t]
	\centering
	\includegraphics{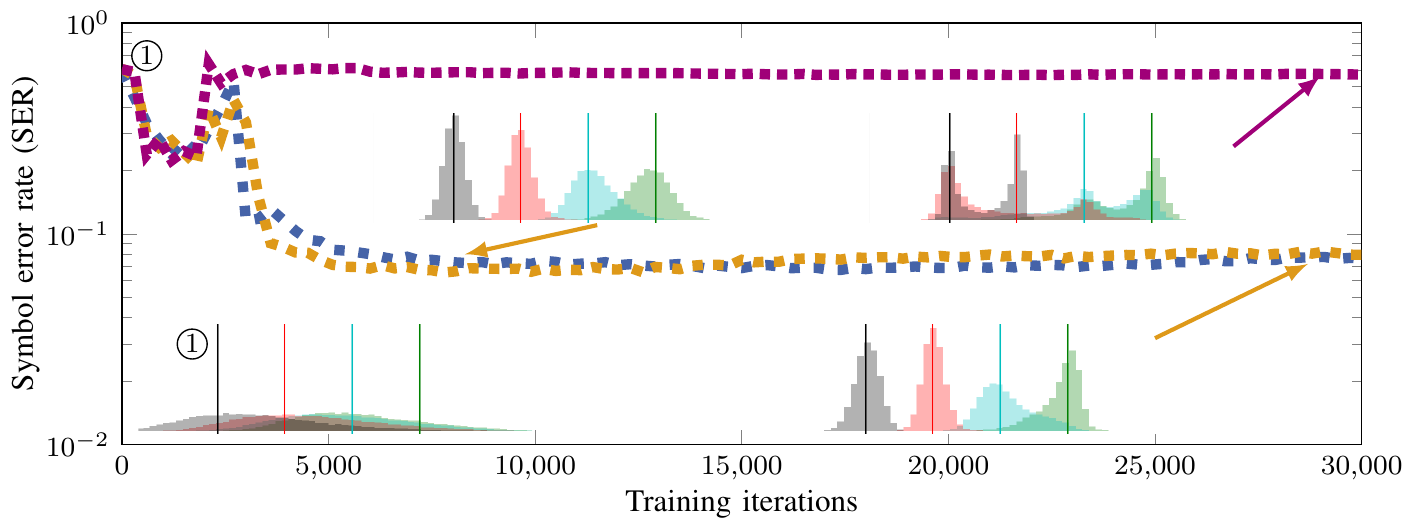}
	\vspace*{-1ex}
	\caption{Convergence behavior of the GAN-CNN for PAM-4 transmission over an \ac{imdd} channel with $R_\mathrm{S}=\SI{25}{\giga\baud}$, $\Lf=\SI{15}{\kilo\meter}$, $\snr=\SI{20}{dB}$, and $\elr|_\mathrm{G}=0.001$, $\elr|_\mathrm{D}=0.001$. Depicted are three training runs and selected histograms of the equalizer output $\bm{z}$ at different training states. The light vertical lines represent the positions of the transmitted symbols with the corresponding color coding.}
	\vspace*{-1ex}
	\label{fig:PAM4_verlauf}
\end{figure*}

The convergence behaviour of three different runs is depicted in Fig.~\ref{fig:PAM4_verlauf} for PAM-4 transmission. Histograms of the generator output $\bm{z}$ highlight the issues during training, i.e., the misclassification of the symbols in the violet (not-converged) curve. Precisely, the transmit symbols are not bijectively projected to a single but to multiple output symbols, e.g., the first (black) transmit symbol is projected to the first, second and third symbol level at the output. From a theoretical perspective, the system optimized its objective $P\mlr{\hat{\bm{x}}} = P\mlr{\bm{x}}$ but violated the requirement $(i)$ of \eqref{IT_requirements}. 

In summary, the presented results demonstrate the high potential of our approach to improve state-of-the-art equalization in terms of flexibility, automation and communication performance. Even though the GAN-based equalizer operates blindly, i.e., without any knowledge of the transmitted symbols, results show that it can converge to---or even outperform---the supervised baselines. This allows to decrease the overhead and to increase the achievable information rate, which is crucial for modern communication systems. 

\section{Conclusion} \label{Sec:conclusion}
In this work, we have proposed a novel approach for blind and channel-agnostic equalization, and proved the concept's feasibility. In simulations of both linear and non-linear channels with different modulation formats, we are able to demonstrate the approach's flexibility and potential to converge towards the performance of non-blind reference equalizers.
We want to highlight that the proposed equalizer is adaptive and can be used for time-varying channels, which is part of our ongoing investigations.
Although challenges like training instability are not yet fully solved, the results and the theoretical analysis show that the novel concept can potentially outperform state-of-the-art equalizers while being blind and channel-agnostic. Thus, the proposed equalizer in this work is highly relevant for future research and is a next step towards integrating deep learning into communication systems.

\end{document}